\documentclass[aps,prl,unsortedaddress,twocolumn,nofootinbib,showpacs,nofootsinbib]{revtex4-1}

\usepackage{amsfonts}

\usepackage[pdftex]{graphicx}
\usepackage{epstopdf}
\usepackage{amsmath}
\usepackage{amssymb}
\usepackage{verbatim}
\usepackage[ulem=normalem]{changes}
\usepackage{booktabs}
\usepackage{xcolor}

\usepackage{soul}

\newcommand{\bra    }{\langle}
\newcommand{\ket    }{\rangle}

\usepackage[T1]{fontenc}

\begin{document}

\title{Decision Making in the Arrow of Time}

\author{\'Edgar Rold\'an$^{1,5}$, Izaak Neri$^{1,2,5}$,  Meik D\"orpinghaus$^{3,5}$, \\ 
Heinrich Meyr$^{3,4,5}$,}
\author{Frank J\"ulicher$^{1,5}$}\email{Corresponding author: julicher@pks.mpg.de}
\medskip

\affiliation{$^1$ Max Planck Institute for the Physics of Complex Systems, N{\"o}thnitzer Str. 38, 01187 Dresden, Germany. \\
$^2$ Max Planck Institute of Molecular Cell Biology and Genetics, Pfotenhauerstra{\ss}e 108, 01307 Dresden, Germany. \\
$^3$Vodafone Chair Mobile Communications Systems, Technische Universit\"at Dresden, 01062 Dresden, Germany. \\
$^4$Institute for Integrated Signal Processing Systems, RWTH Aachen University, 52056 Aachen, Germany.\\
$^5$Center for Advancing Electronics Dresden, cfaed, Germany.}

\begin{abstract}

We show that the steady state entropy production rate of a stochastic process is inversely proportional to the minimal time needed to decide on the direction of the arrow of time. Here we apply Wald's sequential probability ratio test to optimally decide on the direction of time's arrow in stationary Markov processes. Furthermore the steady state entropy production rate can be estimated using mean first-passage times of suitable physical variables. We derive a first-passage time fluctuation theorem which implies that  the decision time distributions for correct and wrong decisions are equal. Our results are illustrated by numerical simulations of two simple examples of nonequilibrium processes.

\end{abstract}


\maketitle

Processes that take place far from thermodynamic equilibrium are in general irreversible and are associated with entropy production. Irreversibility implies that a sequence of events  that takes place during a process occurs with different probability than the same sequence in time-reversed order. Irreversibility and the thermodynamic arrow of time can be illustrated considering a movie displaying the evolution of a complex dynamic process. Such a movie can be run either forward in time or in reverse.   For an irreversible process it is possible to decide  whether the movie is run forward or in reverse defining the direction of the arrow of time by the direction in which entropy increases on average~\cite{jarzynski2011equalities}. For a system at thermodynamic equilibrium, however, even though all atoms or molecules move rapidly in all directions, it is impossible when watching a movie to tell whether it runs forward or in reverse. This raises the following question: Can  the time $\tau_{\rm dec}$ needed  to decide between two hypotheses (movie run forward or in reverse)  be related quantitatively to the degree of irreversibility and the rate of entropy production?

\begin{figure}
\includegraphics[width=8cm]{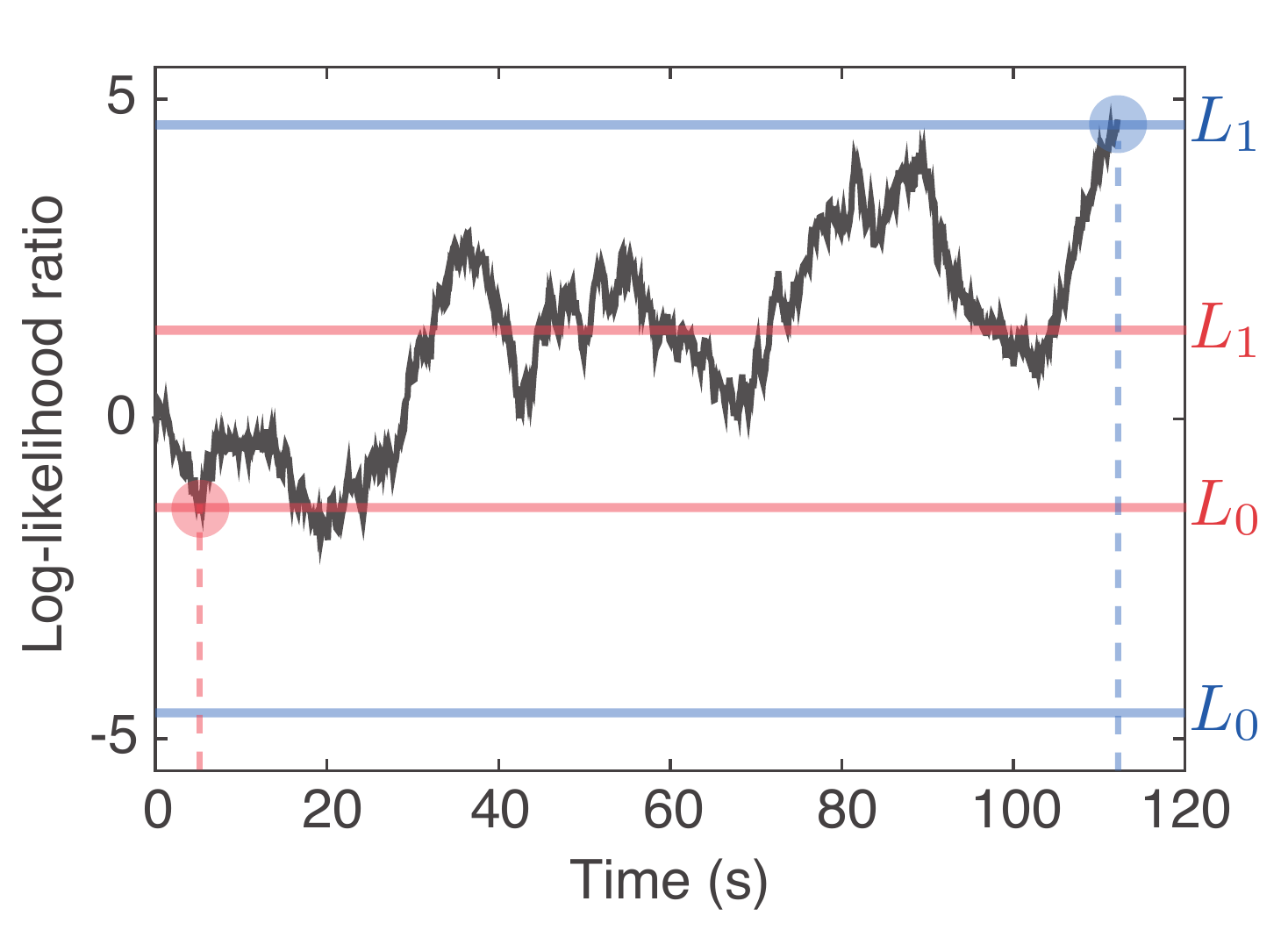} 
\caption{Log-likelihood ratio of the sequential probability ratio test in the arrow of time as a function of time in a drift-diffusion process with diffusion coefficient  $D=0.52 \,\mu\text{m}^2/s$ and drift velocity $v=65\,\mu\text{m}/s$.  The simulation time step is $\Delta t = 0.1\,\rm ms$. The thresholds of the test are shown as horizontal lines for symmetric error probabilities equal to $20\%$  (red) and $1\%$ (blue). The thresholds $L_0$ and $L_1$ correspond to the choice of one of the two hypotheses: the sequence runs forwards ($L_1$) or backwards ($L_0$) in time. With $20\%$ error probability, the decision is made faster (red circle and vertical red dashed line) than for  $1\%$ error probability (blue circle and vertical blue dashed line).\label{fig:waldtest}}
\end{figure}

Decision theory provides a general theoretical framework to optimally make decisions  based on  observations of stochastic processes~\cite{melsa1978decision}. 
An important question of decision theory is what is the earliest time to make a decision $d$ between  two competing hypothesis  $H_1$ and $H_0$ with a given reliability, while  observing a stochastic process. 
In 1943, A. Wald made a pioneering contribution  to this problem by  introducing the sequential  probability ratio test (SPRT)~\cite{wald1945sequential}, which provides the minimal mean decision time for a broad class of stochastic processes~\cite{tartakovsky2014sequential}. 
Wald's SPRT states that the decision $d=1$ ($d=0$)  should be made when the cumulated logarithm of the likelihood ratio $\mathcal{L}(t)$ for the first time exceeds (falls below) a prescribed threshold $L_1$ ($L_0$) (see Fig. \ref{fig:waldtest}). The thresholds $L_1$ and $L_0$  are determined by the   maximally allowed probabilities to make a wrong decision  $\alpha_1=P(d=1|H_0)$ and $\alpha_0=P(d=0|H_1)$. Here, $\alpha_1$ ($\alpha_0$) is the probability to incorrectly make the decision $d=1$  ($d=0$) when the hypothesis $H_0$ ($H_1$) is true. 


In this Letter, we derive a general relation between the average entropy production rate in a nonequilibrium steady state and the mean  time   to decide  whether a stationary stochastic process runs forwards ($H_1$) or backwards in time ($H_0$) using the SPRT. Furthermore, we introduce a fluctuation theorem for the first-passage time probability distribution of the total entropy changes and obtain a fluctuation theorem for the decision time distribution of the SPRT in the arrow of time. Our work  reveals that entropy production can be estimated measuring  first-passage  times  of stationary stochastic processes.

We consider a physical system in a nonequilibrium steady state. We denote by  $X_t  = \{ X(s)\}_{s=0}^t$ a path describing the evolution of a state as a function of time $t$. We denote by $\tilde{X}_t$  the time-reversed path $\tilde{X}_t = \{ X(t-s)\}_{s=0}^t$~\cite{footnote0}. The state of the system is characterized by the path probability $P(X_t)$. The entropy production associated with the path $X_t$ can be defined as \cite{maes2003time}
\begin{equation}
 \Delta S_{\rm tot} [X_t]=k \ln\frac{P(X_t)}{P(\tilde{X}_t)}\quad,
\label{eq:ep}
\end{equation}
where $k$ is Boltzmann's constant. We now perform a SPRT of two hypotheses. Given a path $X_t$, we want to decide whether it corresponds to a forward or time-reversed trajectory 
of the nonequilibrium steady state. We therefore consider the hypothesis $H_{\rightarrow}=H_1$ that  the path runs forward in time with the conditional probability $P(X_t|H_{\rightarrow})=P(X_t)$; and  the hypothesis $H_{\leftarrow}=H_0$ that the dynamics is time reversed, for which $P(X_t|H_{\leftarrow})=P(\tilde{X}_t)$.  Using the SPRT, the decision is made when  the log-likelihood ratio or Turing's weight of evidence~\cite{good1979studies}
\begin{equation}
\mathcal{L} (t)=\ln\frac{P(X_t|H_{\rightarrow})}{P(X_t|H_{\leftarrow})}\quad
\label{eq:WSTNESS}
\end{equation}
 reaches for the first time one of the thresholds $ L_1=L$ and $L_0=-L$, where we have chosen for simplicity a SPRT with symmetric  decision error probabilities $\alpha_0=\alpha_1=\alpha$. When $\mathcal{L}(t)$ is continuous, we have $L=\ln [(1-\alpha)/\alpha]$~\cite{wald1945sequential}. 
 
 The  entropy production and the log-likelihood ratio are related by 
 \begin{equation}
 \mathcal{L}(t) = \frac{\Delta S_{\rm tot}[X_t]}{k}\quad.
 \label{eq:LS}
 \end{equation}
 This provides  a connection between decision theory and stochastic thermodynamics.  Moreover, it allows us to obtain relations between average decision times in the SPRT and the average rate of entropy production.  Applying the SPRT to continuous time Markov processes (See Supplemental Material~\cite{SI} and Ref.~\cite{phatarfod1965sequential}), we show that the mean decision time for a stochastic process with continuous $\mathcal{L}(t)$ is given by
 \begin{equation}
 \langle \tau_{\rm dec}\rangle=\frac{L(1-2\alpha) + \langle \mathcal{L}_{\rm ex}\rangle_{\rm dec}}{\langle \text{d}\mathcal{L}/\text{d}t \rangle}\quad.
 \label{eq:tauLLR}
 \end{equation}
Here  $\langle\dots\rangle$ denotes an ensemble average in steady state.  An average over the ensemble which starts from an initial distribution of states that equals the distribution of states at the decision times is denoted by $\langle\dots\rangle_{\rm dec}$. The excess log-likelihood ratio $\mathcal{L}_{\rm ex}$ is defined as 
\begin{equation}
\mathcal{L}_{\rm ex}=\int_0^\infty \left[\frac{\text{d}\mathcal{L}}{\text{d} t'} - \left\langle\frac{\text{d}\mathcal{L}}{\text{d} t'}\right\rangle \right] \text{d} t'\quad.
\end{equation}
The mean decision time of the SPRT for independent identically distributed ($\text{i.i.d.}$) observations is a special case of Eq.~\eqref{eq:tauLLR} for which $ \langle \mathcal{L}_{\rm ex}\rangle_{\rm dec}=0$. This is because for an i.i.d.\ process  the state distribution is identical to the stationary distribution. 
 
We now apply the theory of decision times to the SPRT
on the arrow of time of a nonequilibrium process. The
relation~\eqref{eq:LS} together with our Eq.~\eqref{eq:tauLLR} describing the average
decision time can be used to express the average entropy
production rate in steady state as
\begin{equation}
\frac{1}{k}\left\langle \frac{\text{d}S_{\rm tot}}{\text{d}t}\right\rangle=\frac{L(1-2\alpha)+ \langle \Delta S_{\rm ex}\rangle_{\rm dec}/k}{\langle\tau_{\rm dec}\rangle}\quad,
\label{eq:meandectime0}
\end{equation}
where
\begin{equation}
\Delta S_{\rm ex}=\int_0^\infty \left[\frac{\text{d}S_{\rm tot}}{\text{d} t'} - \left\langle\frac{\text{d}S_{\rm tot}}{\text{d} t'}\right\rangle \right] \text{d} t'\quad
\end{equation}
denotes the excess total entropy change.


In the limit of small $\alpha$, the mean decision time becomes large, $\langle \Delta S_{\rm ex}\rangle_{\rm dec}/\langle\tau_{\rm dec}\rangle$ becomes small and, thus, Eq.~\eqref{eq:meandectime0} simplifies to 
\begin{equation}
\frac{1}{k}\left\langle \frac{\text{d}S_{\rm tot}}{\text{d}t}\right\rangle\simeq\frac{L(1-2\alpha)}{\langle\tau_{\rm dec}\rangle}\quad.
\label{eq:meandectime}
\end{equation}
Equations (6) and (8) show that the  minimal average time needed to decide whether a process runs forward or backward in time is inversely proportional to the average entropy production rate. Approaching thermodynamic equilibrium,  the mean decision time diverges because  $\mathcal{L}(t)= 0$ in this limit.  If the reliability of the decision is increased, decision times increase correspondingly. Because the average entropy production rate is a property of the process only and not of the SPRT, the ratio given in the right hand side of~\eqref{eq:meandectime} is thus independent of the error probability $\alpha$.


Making decisions in the arrow of time provides a novel way to estimate the entropy production rate of nonequilibrium Markovian processes.  Estimators for the steady state entropy production rate can be obtained from the first-passage times $ \tau$ of a suitable physical observable $\Gamma(X_t)$. If we use the first-passage  of a physical observable through a threshold value to decide on the arrow of time and if this decision has an error probability $\alpha$, then it follows from the optimality of the SPRT that $\langle\tau\rangle\geq \langle{\tau}_{\rm dec}\rangle$, i.e.,  the  mean first-passage time  $\langle \tau \rangle$  is larger or equal to the mean decision time of the SPRT given in Eq. (8). The resulting estimator of the entropy production provides a lower bound to the exact value: 
\begin{equation}
\frac{1}{k}\left\langle \frac{\text{d}S_{\rm tot}}{\text{d}t} \right\rangle  \geq \frac{D[\rightarrow||\leftarrow]}{\langle \tau\rangle}\quad,
\label{eq:bound}  
\end{equation}
where    $D[\rightarrow||\leftarrow] = D[P(d|H_{\rightarrow})||P(d|H_{\leftarrow})]= \ln [(1-\alpha)/\alpha](1-2\alpha)$ is the Kullback-Leibler divergence between the conditional probabilities of the decision variable. When decisions in the direction of time are made based on the log-likelihood ratio, i.e. when $\Gamma = \Delta S_{\rm tot}/k$, Eq.~(9) becomes an equality.

The stochastic nature of decision making in the arrow of time can be characterized by the probability density $P(\tau_{\rm dec})$ of making a decision at time $\tau_{\rm dec}$. The connection between decision theory and thermodynamics implies a relation between the decision time distribution and the distribution of entropy production $\Delta S_{\rm tot}$. For Markovian processes, the probability density $P(\Delta S_{\rm tot};t)$ of entropy production $\Delta S_{\rm tot}$ during the time interval $t$  is related by a fluctuation theorem to the probability density to reduce entropy by the same amount: $P(\Delta S_{\rm tot};t)/P(-\Delta S_{\rm tot};t) = \exp(\Delta S_{\rm tot}/k)$ \cite{evans1993probability,gallavotti1995dynamical,kurchan1998fluctuation,lebowitz1999gallavotti,seifert2005entropy}. In addition, we find  that the probability distribution of the first-passage time $\tau$ of entropy production also  obeys the following detailed fluctuation theorem if the transition probabilities are translationally invariant~\cite{footnote3} (see Supplemental Material~\cite{SI}):
\begin{equation}
\frac{P (\tau;\Delta S_{\rm tot})}{P (\tau;-\Delta S_{\rm tot})} = \exp\,(\Delta S_{\rm tot}/k)\quad.
\label{eq:FTFPT}
\end{equation}
Here, $P (\tau;\Delta S_{\rm tot})\text{d}\tau$ denotes the probability to reach  the value $\Delta S_{\rm tot}$  for the first time in the time interval $[\tau,\tau+\text{d}\tau]$ given that the entropy production has not reached $-\Delta S_{\rm tot}$ before.  

The relation between entropy production and the log-likelihood ratio \eqref{eq:LS} together with the first-passage time fluctuation theorem \eqref{eq:FTFPT} implies for the SPRT in the arrow of time
\begin{equation}
\frac{P(\tau_{\rm dec};L)}{P(\tau_{\rm dec};-L)} = \exp\,(L)\quad.
 \label{eq:wrft}
 \end{equation}
Here,  $P(\tau_{\rm dec};L)$ is the probability distribution of the decision time of the SPRT for a given error rate $\alpha$. $P(\tau_{\rm dec};L)$ is also the distribution of first-passage times to reach the threshold $L$ for the first time without reaching the threshold $-L$ before, given $H_{\rightarrow}$ is true. 
The probability distributions in~\eqref{eq:wrft} are equal to the joint probability densities to make a decision $d\in\{\rightarrow,\leftarrow\}$ at time $\tau_{\rm dec}$, $P(\tau_{\rm dec},\rightarrow)=P(\tau_{\rm dec};L)$, and $P(\tau_{\rm dec},\leftarrow)=P(\tau_{\rm dec};-L)$. Equation~\eqref{eq:wrft} thus implies
\begin{equation}
\frac{P(\tau_{\rm dec},\rightarrow)}{P(\tau_{\rm dec},\leftarrow)} = \exp\,(L)\quad.
 \label{eq:FTAT_un}
 \end{equation}
 From Eq. \eqref{eq:FTAT_un} it follows that $P(d=\rightarrow)/P(d=\leftarrow)= \exp(L)$, consistent with previous results obtained for two-boundary first-passage time processes~\cite{gichman,lindsey1977complete}. Using $P(\tau_{\rm dec},d)=P(\tau_{\rm dec}|d)P(d)$, we then find that the conditional probability densities for the decision time obey
\begin{equation}
P(\tau_{\rm dec}|\!\rightarrow) = P(\tau_{\rm dec}|\!\leftarrow) \quad .
\label{eq:FTAT}
\end{equation}
This implies that even though  decisions are made with different probabilities, the conditional decision time distributions have the same shape for both outcomes.  We therefore call Eq. \eqref{eq:FTAT} the Fluctuation Theorem in the Arrow of Time (FTAT). Equations~\eqref{eq:meandectime} and~\eqref{eq:FTAT} are the main results of this paper.

\begin{figure}
\includegraphics[width=6cm]{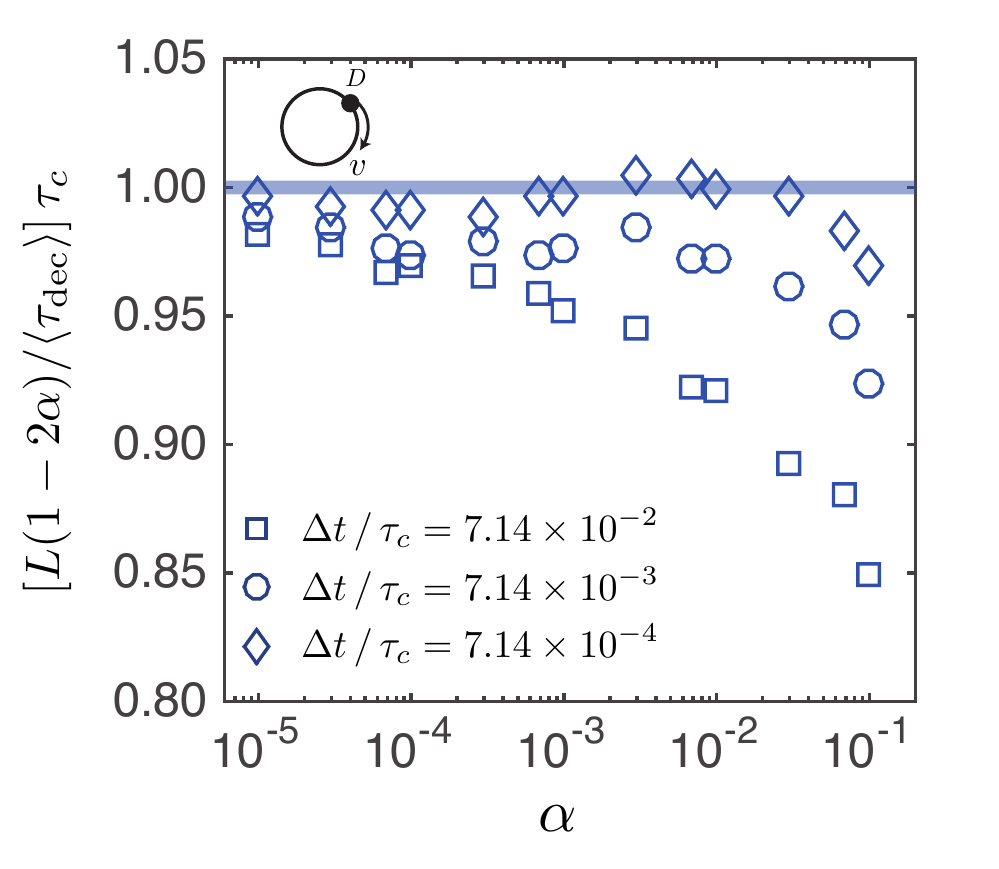} 
\caption{Estimation of the steady state entropy production rate of a drift-diffusion process with periodic boundary conditions (e.g., particle in a ring, see inset) by a sequential probability ratio test in the arrow of time. The estimator $L(1-2\alpha)/\langle\tau_{\rm dec}\rangle$ is shown   as a function of the error probability  $\alpha$ for different simulation time steps $\Delta t /\tau_c$ and normalized by $\tau_c=1/[\langle \text{d}S_{\rm tot} /\text{d}t\rangle / k]=D/v^2$. For the vertical axis we use the empirical mean of $\tau_{\rm dec}$ and $L$ is the threshold for the decision. 
 The data is  obtained from $1000$ numerical simulations with drift velocity $v=65\,\mu\text{m}/s$ and diffusion coefficient $D=0.52 \,\mu\text{m}^2/s$. The horizontal line corresponds to the steady state entropy production rate.
\label{fig:tau_S_F}}
\end{figure}

To illustrate how Eq.~\eqref{eq:meandectime} provides an estimator for the steady state entropy production rate, we discuss two paradigmatic examples of nonequilibrium stochastic processes. We first consider a drift-diffusion process with periodic boundary conditions of a particle with position $x(t)$, average drift velocity $v$, and diffusion coefficient $D$. If Einstein's relation holds, $D=kT/\gamma$ where $\gamma$ is a friction coefficient, the steady state entropy production rate  is $\langle \text{d}S_{\rm tot} /\text{d}t\rangle / k= v^2/D= F^2/(\gamma k T)$, where  $F=\gamma v$ is the friction force and $T$ is the temperature of the thermal bath~\cite{seifert2012stochastic}.  Figure \ref{fig:tau_S_F}
shows $L(1-2\alpha)/\langle\tau_{\rm dec}\rangle$ obtained from $1000$ numerical simulations of the SPRT in the arrow of time (markers) as a function of the error probability $\alpha$ for different values of the simulation time step $\Delta t$ together with $\langle \text{d}S_{\rm tot} /\text{d}t\rangle/k $ (blue solid line).  For the drift-diffusion process, the log-likelihood ratio for the SPRT in the arrow of time is simply given by $\mathcal{L} (t)= (v/D) [x(t) - x(0)]$. As long as the simulation time step obeys $\Delta t\ll  \tau_c $, where $\tau_c = k/\langle \text{d}S_{\rm tot} /\text{d}t\rangle = D/v^2$, the SPRT in the arrow of time provides an accurate estimator of entropy production independent of the error probability $\alpha$. For larger $\Delta t$, the estimator is  only accurate for small $\alpha$ but provides a lower bound to the steady state entropy production rate for larger $\alpha$. In our simulations, we also calculated empirical conditional decision time probabilities $P(\tau_{\rm dec}|\!\rightarrow)$ (green) and $P(\tau_{\rm dec}|\!\leftarrow)$ (purple), which are shown in Fig. 3 for $\alpha = 0.01$. Figure 3 demonstrates the validity of the FTAT given in \eqref{eq:FTAT} for  the drift-diffusion process.

\begin{figure}
\includegraphics[width=6cm]{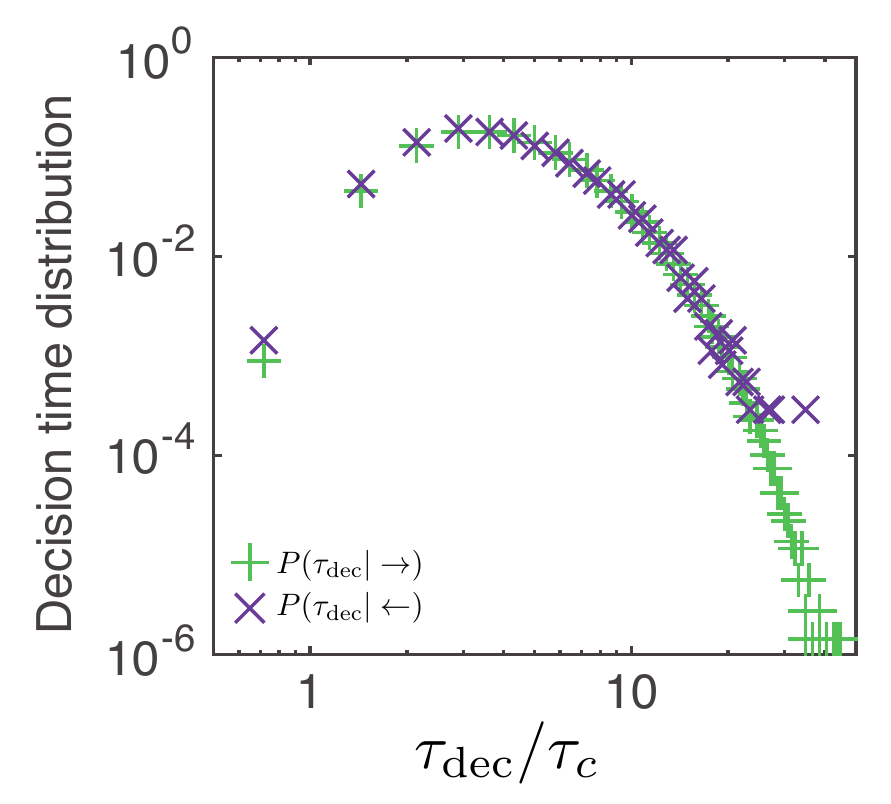} 
\caption{Conditional distributions of the decision time $P(\tau_{\rm dec}|\!\rightarrow)$  and $P(\tau_{\rm dec}|\!\leftarrow)$  obtained from $10^6$ numerical simulations of a drift diffusion process with diffusion $D=0.52 \,\mu\text{m}^2/s$ and drift $v=65\,\mu\text{m}/s$. The simulation time step is $\Delta t = 0.1 \,\rm ms$ and the error probability $\alpha = 0.01$. \label{fig:FT-FPT}}
\end{figure}

\begin{figure}
\includegraphics[width=6.5cm]{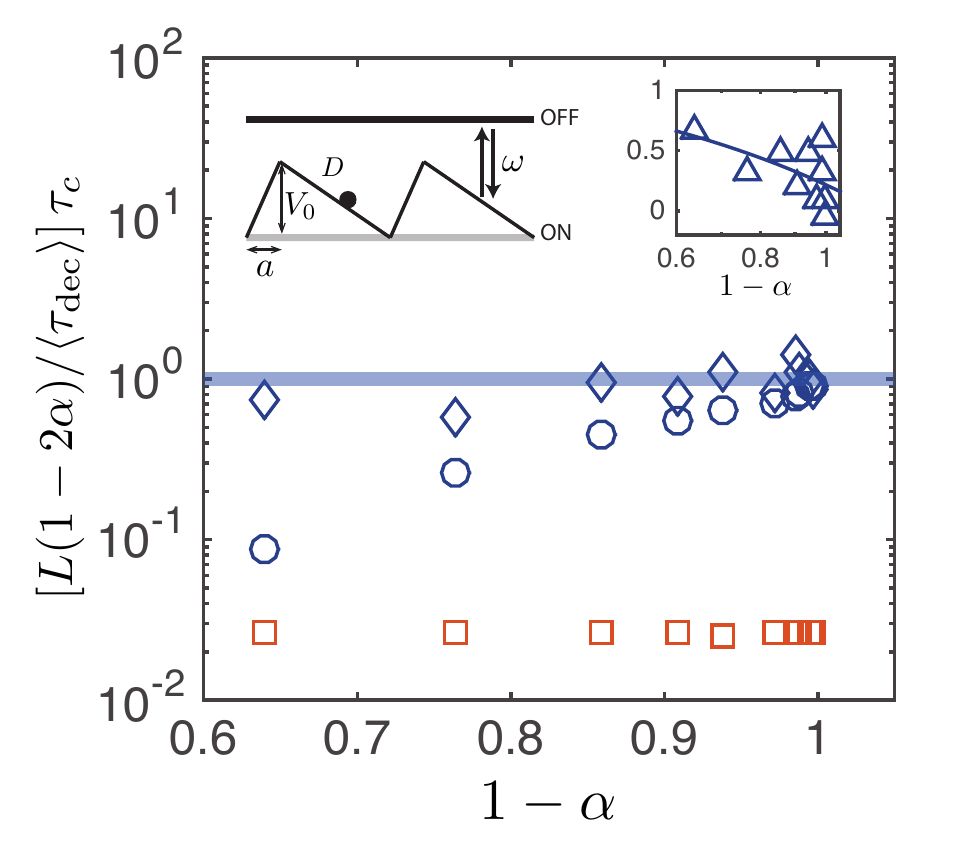} 
\caption{Estimation of the entropy production rate using the SPRT in the arrow of time in a flashing ratchet model using the right hand side in Eq. \eqref{eq:meandectime} (blue open circles) and the right hand side in Eq. \eqref{eq:meandectime0} (blue open diamonds) as a function of the reliability of the test. Red open squares are given by the ratio between $D[\rightarrow||\leftarrow]$ and the mean first-passage time $\langle\tau \rangle$ of the position of the particle  in Eq.~\eqref{eq:bound}. 
The results were obtained from $1000$ numerical simulations with time step $\Delta t = 1\,\rm \mu s$, diffusion coefficient $D=1\,\mu\text{m}^2/\text{ms}$, $V_0=10 \,kT$, $a=1/3 \,\mu\text{m}$, and $\omega = 10 \,\text{kHz}$. 
The characteristic time $\tau_c=0.07\,\rm ms$ is the numerical estimate of $k/\langle \text{d}S_{\rm tot}/\text{d}t\rangle$ obtained from a single stationary trajectory of $10^7$ data points.
Inset: Correction term in  $\eqref{eq:meandectime0}$ given by $[\langle \Delta S_{\rm ex}\rangle_{\rm dec}/k {\langle\tau_{\rm dec}\rangle}]\,\tau_c$ as a function of $1-\alpha$ (blue triangles). The solid line is a linear fit of the data.
\label{fig:ratchet}}
\end{figure}

\begin{figure}
\includegraphics[width=8.5cm]{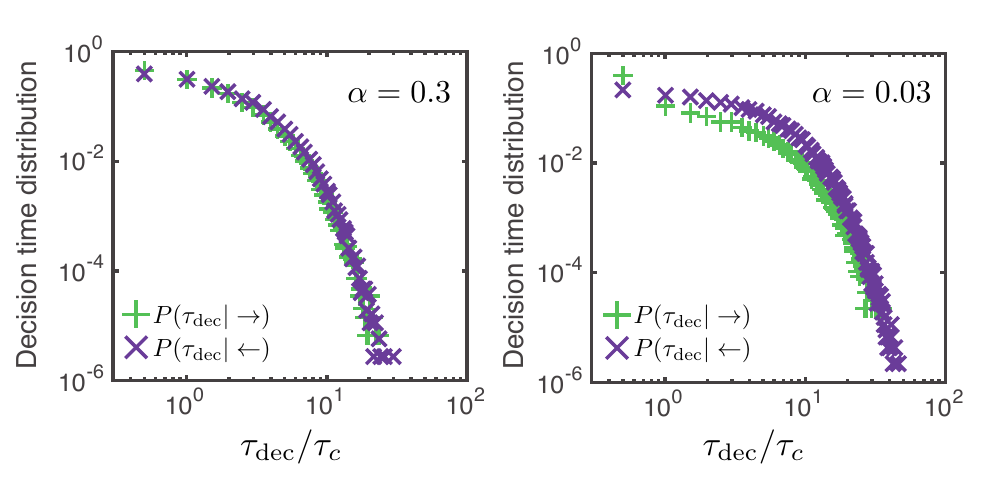} 
\caption{Conditional distributions of the decision time $P(\tau_{\rm dec}|\!\rightarrow)$ and $P(\tau_{\rm dec}|\!\leftarrow)$  obtained from $10^6$ numerical simulations of the flashing ratchet with the same simulation parameters as in Fig.~\ref{fig:ratchet}. The two figures show the distributions for two different error probabilities $\alpha$. \label{fig:ftratchet}}
\end{figure}

The drift-diffusion process is a simple example and serves as an illustration of our results. We now test whether our results also hold in  more complex nonequilibrium stochastic processes that involve discontinuous jumps of the state variables. We therefore discuss the SPRT in the arrow of time for the case of a flashing ratchet with periodic boundary conditions. We consider a Brownian particle with diffusion coefficient $D$, subject to a piecewise linear periodic potential that is  switched on and off stochastically at a constant rate $\omega$~\cite{prost1994asymmetric}. The log-likelihood ratio of the SPRT in the arrow of time in steady state can be approximated by the cumulated work  $W$ exerted on the particle during switches $\mathcal{L}(t)=W(t)/kT$. Here $W(t)=\sum_i \Delta V_i$, where $\Delta V_i$ is the potential energy change during the switching event $i$ and the sum is done over all switches that occur before time $t$~\cite{de2007symmetries}.  Figure \ref{fig:ratchet} shows the estimate of  $L(1-2\alpha)/\langle \tau_{\rm dec}\rangle$  as a function of the reliability $1-\alpha$. The plot shows that the SPRT in the arrow of time provides a lower bound for the steady state entropy production rate (blue open circles) and converges for high reliability to the correct value. In addition, Fig.~\ref{fig:ftratchet} shows the conditional distributions of the decision times revealing that the  FTAT holds to good approximation for high error probabilities despite that the propagator is not translationally invariant.

When using the  estimator given by Eq.~\eqref{eq:meandectime0}, which includes the excess entropy production, the entropy production rate is  estimated more accurately  at low reliabilities (Fig. \ref{fig:ratchet}, blue diamonds). The inset in Fig.~\ref{fig:ratchet} confirms that the correction term in Eq.~\eqref{eq:meandectime0} tends to zero for $\alpha$ small. 
 Note that the estimator $L(1-2\alpha)/\langle\tau_{\rm dec}\rangle$ in Eq.~\eqref{eq:meandectime} provides a lower bound at small $\alpha$  because of the discontinuous jumps in the state variables. Using an heuristic estimator given by the ratio $D[\rightarrow||\leftarrow]/\langle \tau\rangle$, where $\tau$  is the first-passage time of the position  of the particle, also bounds from below the steady state entropy production,  as follows from Eq.~\eqref{eq:bound} (Fig.~\ref{fig:ratchet}, red squares).

The dynamics of a stochastic nonequilibrium process provides evidence on the arrow of time to an observer. Reliable decisions on the direction of the arrow of time can be made measuring first-passage times of physical observables. 
 When the physical observable used is the entropy production, the decision time is minimized.   
 In addition, measuring first-passage times of physical observables provides estimators for the steady state entropy production rate that are lower bounds to the true value. This follows from the optimality of the SPRT.   Using this method to estimate entropy production, it is not necessary to 
sample  the  whole space of stochastic trajectories  as required in previous approaches~\cite{andrieux2007entropy,roldan2014irreversibility,tusch2014energy}.  Interestingly, our fluctuation theorem for the two-boundary first-passage time distribution of entropy production~\eqref{eq:FTFPT} implies that the shape of the distributions of decision times for correct  and wrong decisions are equal even though the probabilities in both cases are different.   The connection between decision theory and thermodynamics provided here could be of particular interest in the context of  nonequilibrium processes that involve feedback control, often found  in biology and engineering. 

We acknowledge fruitful discussions with Mostafa Khalili-Marandi. This work is partly supported by the German Research Foundation (DFG) within the Cluster of Excellence 'Center for Advancing Electronics Dresden'. ER acknowledges funding from Grupo Interdisciplinar de Sistemas Complejos (GISC, Madrid, Spain) and from Spanish Government, grants ENFASIS (FIS2011-22644) and TerMic (FIS2014-52486-R).

%

\newpage
\section*{Supplemental Material}
\section{S1. Mean decision time of the SPRT in continuous-time stationary Markov processes}

The aim of this section is to show the derivation of Eq. (4) in the Main Text. We first review the SPRT for Markov chains (Sec.~S1A) and the corresponding mean decision time formula obtained in Ref.~\cite{phatarfod1965sequential} (Sec.~S1B). Using these results we then derive Eq. (4) in the Main Text relating the mean decision time to the rate of change of the log-likelihood ratio for the SPRT between continuous time Markov processes (Sec.~S1C). Finally we provide a full derivation  of Eq.~\eqref{eq:eqap} to help the reader (Sec.~S1D).

\subsection{A. Sequential probability ratio test for stationary Markov chains}
\label{sec:SPRTM}
We consider a discrete-time sequence  $X_1^n =(x_1, x_2, \cdots, x_n)$ where the variables $x_i$ are elements of a countable set of observations $\Omega$. The sequence is described by a stationary Markov process and therefore $P(X_1^n)=P_{\rm{ss}}(x_1)P(x_2|x_1)\cdots P(x_n|x_{n-1})$, where $P_{\rm ss}(\cdot)$ is the stationary distribution and $P(x_j|x_i)$ is the transition probability from $x_i$ to $x_j$. Based on the sequence $X_1^n$, we consider two competing hypotheses $H_0$ and $H_1$ for which $P(X_1^n|H_0)=P_{\rm ss}(x_1) P_0(x_2|x_1)\cdots P_0(x_n|x_{n-1})$ and $P(X_1^n|H_1)=P_{\rm ss}(x_1) P_1(x_2|x_1)\cdots P_1(x_n|x_{n-1})$, respectively. Here we consider the case where the stationary distribution is the same under both hypotheses. The sequential probability ratio test (SPRT) between $H_0$ and $H_1$  follows the evolution in time of the following log-likelihood ratio $\mathcal{L}_n$
\begin{eqnarray}
 \mathcal{L}_n &=& \ln\frac{P (X_1^n|H_1) }{P (X_1^n|H_0) }\quad, \\ 
 &=&\sum_{i=1}^{n-1} \ln \frac{P_1(x_{i+1}|x_{i})}{P_0(x_{i+1}|x_{i})}\quad. \label{eq:LLR}
\end{eqnarray}
When $L_0<\mathcal{L}_n<L_1$ the test continues and a new observation $x_{n+1}$ is taken.   When $\mathcal{L}_n\geq L_1$ the test terminates and the hypothesis $H_1$ is accepted and when  $\mathcal{L}_n \leq L_0$  the test terminates and the hypothesis $H_0$ is accepted.  The threshold values are approximatively given by
\begin{eqnarray}
 L_1 = \ln\left(\frac{1-\alpha_1}{\alpha_0}\right)\quad,\\
 L_0 = \ln\left(\frac{\alpha_1}{1-\alpha_0}\right)\quad,
\end{eqnarray}
with $\alpha_0$ the allowed error probability to decide for $H_1$, when the hypothesis $H_0$ is true, and  $\alpha_1$ the allowed error probability to decide for $H_0$, when the hypothesis $H_1$ is true. For a continuous-time sequence of observations the threshold values are exact when the log-likelihood ratio is a continuous function in time~\cite{wald1945sequential}.

One can equivalently write the transition matrices in Dirac's bra-ket notation as follows: $P(x_j|x_i)=\bra j|\mathsf{P}|i\ket$,  $P_0(x_j|x_i)=\bra j|\mathsf{P}_0|i\ket$  and $P_1(x_j|x_i)=\bra j|\mathsf{P}_1|i\ket$. Similarly, for the stationary distribution $P_{\rm ss}(x_i)= \bra i |  p_{\rm ss}\ket$.

\subsection{B. Mean number of observations in the SPRT for Markov chains}
\label{sec:phatarfod}

Given a sequence of observations, $n_{\rm dec}$ is defined as the number of observations taken until the test terminates. The mean number of observations $\mathbb{E}[ n_{\rm dec} ]$ to make a decision (for either $H_0$ or $H_1$) over all possible sequences is, following Phatarfod's derivation, given by~\cite{phatarfod1965sequential} 
\begin{eqnarray}
\mathbb{E}[ n_{\rm dec} ]= \frac{\mathbb{E}[  \mathcal{L}_{n_{\rm dec}}]    - \bra l'_1(0)|p_{\rm ss}\ket + \mathbb{E} \left[ \bra l'_1(0)|x_{n_{\rm dec}} \ket \right] }{\lambda'_1(0)}\;.
\label{eq:meandec1}
\end{eqnarray}
Note that in the corresponding equation (3.10) in Ref.~\cite{phatarfod1965sequential} there is a typo and the sign of the last two terms in the numerator are swapped. Here $\mathbb{E}$ denotes an average over all possible infinitely long sequences\footnote{The expectation $\mathbb{E}[\cdot]$ corresponds to the ensemble average $\langle\cdot\rangle$ in the Main Text. This is because we use the bra-ket notation for vectors and matrices.}. The first term in the numerator $\mathbb{E}[  \mathcal{L}_{n_{\rm dec}}]$ is the expected value of the log-likelihood ratio when the decision is taken. The rest of the terms can be defined in terms of the following generating matrix,
\begin{eqnarray}
 \mathbf{P}(z) = \sum_{i,j\in \Omega}|i\ket\bra j |  \:\bra i| \mathsf{P} |j \ket\: \exp\left[{z\ln\left(\frac{\bra i|\mathsf{P}_{1}| j\ket}{\bra i|\mathsf{P}_{0}|j\ket}\right)}\right]\;,
 \label{eq:Pz}
\end{eqnarray}
where $z\in\mathbb{R}$. The matrix $ \mathbf{P}(z)$ has eigenvalues $\lambda_i(z)$  with corresponding right eigenvectors $|r_i(z) \ket$ and left eigenvectors $\bra  l_i(z)|$.  Here $\lambda_1(0)=1$ is the Perron root of the transition matrix $\mathsf{P}$ and $\lambda_1(z)$ is the corresponding  eigenvalue of $ \mathbf{P}(z) $.  Thus $| r_1(z)\ket$ is the right eigenvector of $\lambda_1(z)$ and $\bra l_1(z)|$ is the corresponding left eigenvector. In the limit of $z\to 0$, $| r_1(0)\ket = | p_{\rm ss}\ket$ and $\bra l_1(0)|= \bra 1|$ for which $ \bra 1|i\ket=1$ for all $i\in\Omega$. The denominator in Eq.~\eqref{eq:meandec1} equals $\lambda'_1(0)=\lim_{z\to 0}\frac{\text{d}}{\text{d}z} \lambda_1(z)$ and $\bra l_1'(0)|= \lim_{z\to 0}\frac{\text{d}}{\text{d}z}\bra l_1(z)|$. The remaining term $\mathbb{E} \left[ \bra l'_1(0)|x_{n_{\rm dec}} \ket \right] $ contains the ensemble average of the right eigenvector $| x_{n_{\rm dec}}\ket$ corresponding to the value of the observation $x_{n_{\rm dec}}$ when the decision is taken: $\bra j |  x_{n_{\rm dec}}\ket = \delta_{j,x_{n_{\rm dec}}}$, where $\delta$ denotes the Kronecker delta.

\subsection{C. Mean decision time formula in the continuous time limit}
\label{sec:explicit}

We first revisit matrix perturbation theory and then apply it to determine the different terms in Phatarfod's mean decision time formula given by Eq.~\eqref{eq:meandec1}.

\subsubsection{1. Revision of matrix perturbation theory}

We will use some results from perturbation theory, see Chapter 2 of~\cite{wilkinsonalgebraic}.  
For two matrices $\mathsf{A}$ and $\mathsf{B}$ which satisfy the relations $\left|\bra i|\mathsf{A}|j\ket\right|<1$ and $\left|\bra i|\mathsf{B}|j\ket\right|<1$  we consider the perturbed matrix $\mathsf{C}(z)$
\begin{eqnarray}
 \mathsf{C}(z) = \mathsf{A} + z\:\mathsf{B}\quad,
\end{eqnarray}
with $z\in \mathbb{R}$ the perturbation parameter. The simple eigenvalues $\lambda_k(z)$ of $\mathsf{C}(z)$, and their corresponding eigenvectors $|r_k(z)\ket$ and $\bra l_k(z)|$  are given by a convergent power series in the perturbation parameter $z$
\begin{eqnarray}
 \lambda_k(z) &=& \lambda_k (0) + z\: \lambda'_k (0)+ \frac{z^2}{2}\:  \lambda''_k (0) + \cdots  \\ 
|r_k(z)\ket &=& |r_k(0)\ket  + z\:|r'_k(0)\ket  + \frac{z^2}{2}\: |r''_k(0)\ket + \cdots  \\
\bra l_k(z)| &=& \bra l_k(0)| + z\:\bra l'_k(0)| + \frac{z^2}{2}\:  \bra l''_k(0)| + \cdots 
\end{eqnarray}
for sufficient small values of $|z|$. The first order terms of the power series are given by~\cite{wilkinsonalgebraic}
\begin{eqnarray}
  \lambda'_k (0) &=& \frac{\bra l_k(0)| \mathsf{B} |r_k(0)\ket}{\bra  l_k(0)|r_k(0)\ket} \label{eq:lambdapert}\\
|r'_k(0)\ket &=&  \sum_{i\in \Omega\setminus \{k\}}\frac{|r_i(0)\ket\bra l_i(0)| \mathsf{B} |r_k(0)\ket  }{(\lambda_k(0)-\lambda_i(0))\bra l_i (0)|r_i(0)\ket}  \\ 
\bra l'_k(0)| &=& \sum_{i\in \Omega\setminus \{k\}}\frac{\bra l_k(0)| \mathsf{B} |r_i(0)\ket \bra l_i(0)|  }{(\lambda_k(0)-\lambda_i(0))\bra l_i(0)|r_i(0)\ket} \label{eq:lpert}
\end{eqnarray}

\subsubsection{2. Calculation of the mean decision time }

The expansion of the exponential in Eq.~\eqref{eq:Pz} for $z$ small yields
\begin{eqnarray}
 \mathbf{P}(z) = \mathsf{P} + z\:\mathsf{P}' + \mathcal{O}(z^2) \quad.
\end{eqnarray}
with $\mathsf{P}$ the transition probability matrix and 
\begin{equation}
\mathsf{P}'= \sum_{i,j\in\Omega}|i\ket   \bra i|\mathsf{P}|j\ket \bra j| \;\ln\left(\frac{\bra i|\mathsf{P}_{1}|j\ket}{\bra i|\mathsf{P}_{0}|j\ket}\right) \quad.
\end{equation}
Identifying in Eqs.~\eqref{eq:lambdapert} and~\eqref{eq:lpert} $\mathsf{C}(z)$ as $\mathbf{P}(z)$, $\mathsf{A}$ as $\mathsf{P}$ and $\mathsf{B}$ as $\mathsf{P}'$, we obtain
\begin{eqnarray}
 \lambda_1'(0) =  \sum_{i,j\in\Omega}\bra i|\mathsf{P}|j\ket  \bra j|p_{\rm ss}\ket \:\ln\left(\frac{\bra i|\mathsf{P}_{1}|j\ket}{\bra i|\mathsf{P}_{0}|j\ket}\right)\quad.
\end{eqnarray}
We further have (see Appendix)
\begin{eqnarray}
   \bra l'_1(0)|p \ket &=&  \sum_{i,j\in\Omega} \bra i|\mathsf{P}|j\ket \ln\left(\frac{\bra i|\mathsf{P}_{1}|j\ket}{\bra i|\mathsf{P}_{0}|j\ket}\right)  \nonumber\\ 
&\times &\left[\sum^\infty_{n=0}(\bra j |p_n\ket  -\bra j |p_{\rm ss}\ket)\right] \label{eq:eqap}  
\label{eq:exc}
\end{eqnarray}
with  $|p_n\ket  = \mathsf{P}^n |p\ket$  with $|p\ket$ a distribution on phase space or in other words a normalized right eigenvector with $\bra 1 | p\ket = 1$.  For $|p\ket=|p_{\rm ss}\ket$, we thus have $  \bra l'_1(0)|p_{\rm ss} \ket=0$, and therefore the second term in the numerator in Eq.~\eqref{eq:meandec1} vanishes.

Using the results shown above we can rewrite Phatarfod's mean decision time formula~\eqref{eq:meandec1} as 
\begin{eqnarray}
   \mathbb{E} \left[ n_{\rm dec}\right]= \frac{ \gamma L_0 + (1-\gamma)L_1 + \mathbb{E}_{\rm dec}[ \mathcal{L}_{\rm ex}]   }{
\sum_{i,j\in\Omega}\bra i|\mathsf{P}|j\ket \bra j|p_{\rm ss}\ket \:\ln\left(\frac{\bra i|\mathsf{P}_{1}|j\ket}{\bra i|\mathsf{P}_{0}|j\ket}\right)}\label{eq:meandec2}
\end{eqnarray}
where $\gamma$ is the probability to hit $L_0$. For deriving Eq.~\eqref{eq:meandec2} we have used that the log-likelihood ratio at the decision $\mathcal{L}_{n_{\rm dec}}$ reaches exactly one of the two values $L_0$ or $L_1$, and therefore $\mathbb{E}[  \mathcal{L}_{n_{\rm dec}}]  =  \gamma L_0 + (1-\gamma)L_1$. We have also defined the expected value of the excess log-likelihood ratio starting from the state at the decision time
\begin{eqnarray}
\mathbb{E}_{\rm dec}[\mathcal{L}_{\rm ex}] &=&\sum_{i,j\in\Omega} \bra i|\mathsf{P}|j\ket \ln\left(\frac{\bra i|\mathsf{P}_{1}|j\ket}{\bra i|\mathsf{P}_{0}|j\ket}\right)  \nonumber \\
&\times& \left[\sum^\infty_{n=0}(\bra j | \mathsf{P}^n |p_{\rm dec}\ket  -\bra j |p_{\rm ss}\ket)\right],
\end{eqnarray}
with $p_{\rm dec}= \mathbb{E} [ x_{n_{\rm dec}} ]$.
It is now possible to take the continuous time limit of the log-likelihood ratio in Eq.~\eqref{eq:LLR} which yields 
for Eq.~\eqref{eq:meandec2}
\begin{equation}
\mathbb{E}[ \tau_{\rm dec} ]=\frac{\gamma L_0 + (1-\gamma) L_1 + \mathbb{E}_{\rm dec}[ \mathcal{L}_{\rm ex}]}{ \mathbb{E}[ \text{d}\mathcal{L}/\text{d}t ]}\quad,\label{eq:meandectime3}
\end{equation}
where $\text{d}\mathcal{L}(t)/\text{d}t= \lim_{\Delta t \to 0} (1/\Delta t) [\mathcal{L}(t+\Delta t)-\mathcal{L}(t)]$. For symmetric error probabilities, $\alpha_0=\alpha_1=\alpha$, $L_0=-L_1=-L$ and therefore $\gamma L_0 + (1-\gamma) L_1 = L(1-2\alpha)$. In this case, Eq.~\eqref{eq:meandectime3} corresponds to Eq. (4) in the Main Text.

\subsection{D. Derivation of Eq.~(\ref{eq:eqap})}
\label{sec:app}
Let us rewrite Eq.~\eqref{eq:lpert} identifying $\mathsf{C}(z)$ as $\mathbf{P}(z)$, $\mathsf{A}$ as $\mathsf{P}$ and $\mathsf{B}$ as $\mathsf{P}'$,
\begin{eqnarray}
\bra l'_1(0)|p\ket &=& \sum_{i\in \Omega\setminus \{1\}}\frac{\bra 1| \mathsf{P}'|r_i\ket \bra l_i|p\ket  }{(1-\lambda_i)\bra l_i|r_i\ket} 
 \nonumber\\ 
 &=& \sum_{m,j\in\Omega} \bra j|\mathsf{P}|m\ket \ln\left(\frac{\bra j|\mathsf{P}_{1}|m\ket}{\bra j|\mathsf{P}_{0}|m\ket}\right)  \bra m |\mathsf{Q}\: p\ket \;,\nonumber \\ \label{eq:l10p}
\end{eqnarray}
where we have defined the matrix $\mathsf{Q}$ as
\begin{eqnarray}
\mathsf{Q} =  \sum_{i\in \Omega\setminus \{1\}}\frac{|r_i\ket \bra l_i|}{\bra l_i|r_i\ket} (1-\lambda_i)^{-1} \quad, \label{eq:Q0}
\end{eqnarray}
and for simplicity we have used the notation $\lambda_i = \lambda_i (0)$, $|r_i\ket = |r_i (0)\ket$ and $\bra l_i| = \bra l_i (0)|$. Since $\mathsf{P}$ is a transition probability matrix, its eigenvalues $i\neq 1$ satisfy $|\lambda_i|<1$, and therefore we can write
\begin{eqnarray}
 (1-\lambda_i)^{-1} = \sum^\infty_{m=0}\lambda^m_i \quad.
\end{eqnarray}
As a result, 
\begin{eqnarray}
\mathsf{Q} =  \sum^\infty_{m=0} \sum_{i\in \Omega\setminus \{1\}}\frac{|r_i\ket \lambda^m_i \bra l_i|}{\bra l_i|r_i\ket} \quad.
\end{eqnarray} 
which thus is: 
\begin{eqnarray}
\mathsf{Q} = \sum^\infty_{m=0}\left(\mathsf{P}^m-|r_1\ket \bra l_{1}|\right)   \\ 
 = \sum^\infty_{m=0}\left(\mathsf{P}^m-|p_{\rm ss}\ket \bra l_{1}|\right)   \label{eq:Qeq}
\end{eqnarray}
Substituting Eq.~\eqref{eq:Qeq} in~\eqref{eq:l10p} yields Eq.~\eqref{eq:eqap}.

\section{S2. Fluctuation Theorems for First-Passage Time Distributions}

Here we derive the  fluctuation theorem for the distribution of the first passage times of entropy production given by Eq. (10) in the Main Text.

For a translationally invariant and steady state Markovian process $X$, the transition probability $P(x|x' ;  t,t')$  denotes the probability density of $X(t)$ given that $X(t')=x'$. The transition probability satisfies $P(x|x' ;  t,t') = P(x-x'|0;t-t',0)= P(x-x';t-t')$.  Let us assume that the transition probability fulfils a standard fluctuation relation~\cite{gallavotti1995dynamical,kurchan1998fluctuation,lebowitz1999gallavotti,seifert2005entropy},
\begin{eqnarray}
 \frac{P(x;t)}{P(-x;t)} = e^{\eta x}\quad,
 \label{eq:FT}
\end{eqnarray}
where $\eta$ is a positive real number. We now define three different first-passage time distributions: 
\begin{enumerate}
\item The probability $\Psi (t;x)\text{d}t$ that $X$  passes for the first time through a single boundary located at a relative coordinate $x$ in the time interval $[t,t+\text{d}t]$. 
\item The probability $\Phi(t;x)\text{d}t$ that $X$ passes for the first time $x$ in the time interval $[t,t+\text{d}t]$ without having reached $-x$ before. Analogously, $\Phi (t;-x)\text{d}t$ is the probability that $X$ reaches $-x$ for the first time without having reached a barrier located at $x$ before.
\item The probability $\Omega(t;x)\text{d}t$ that $X$ passes for the first time through $x$ in the time interval $[t,t+\text{d}t]$ after having crossed $-x$ at least once  before.
\end{enumerate}
The event, that $X$ reaches unconditionally the single barrier at $x$ equals the probability of the two disjoint events that $X$ has crossed the boundary at $-x$ or has never crossed the boundary at $-x$:
\begin{equation}
\Psi (t;x) = \Phi(t;x) +\Omega(t;x)\quad.
\label{eq:decomposition}
\end{equation}

We first derive a fluctuation relation for the one-boundary first-passage time distribution $\Psi(t;x)$. $\Psi (t;x)$ satisfies the following fluctuation relation
\begin{eqnarray}
 \frac{\Psi (t;x)}{\Psi (t;-x)} = e^{\eta x}\quad. \label{eq:fp}
 \label{eq:propfpt}
\end{eqnarray}
As we will show in the following, Eq.~\eqref{eq:propfpt} follows from the relationship between the transition probability and the one boundary first-passage time distribution~\cite{redner2001guide},
\begin{eqnarray}
 P(x;t) = \int^t_{0} {\rm d}t'\:\Psi(t';x)P(0;t-t')\quad,
\end{eqnarray}
and the Laplace transform of the transition probability equals to
\begin{eqnarray}
 \tilde{P}(x;s) = \tilde{\Psi}(s;x)\tilde{P}(0;s)  \label{eq:Lapl}\quad.
\end{eqnarray}
Here we have introduced the Laplace transforms
\begin{eqnarray}
 \tilde{P}(x;s) &=& \int^\infty_0 {\rm d}t\: e^{-st}P(x;t)\quad, \\ 
 \tilde{\Psi}(s;x) &=& \int^\infty_0 {\rm d}t\: e^{-st}\Psi (t;x)\quad. 
\end{eqnarray}
Since the fluctuation relation is independent of time, the Laplace transform of the transition probability satisfies 
\begin{eqnarray}
 \frac{\tilde{P}(x;s)}{ \tilde{P}(-x;s)} = e^{\eta x}\quad,
 \label{eq:ftlap}
\end{eqnarray}
which together with the factorization property given by Eq.~(\ref{eq:Lapl}) yields a fluctuation theorem for the Laplace transform of the first-passage time distribution $\Psi$:
\begin{eqnarray}
 \frac{\tilde{\Psi}(s;x)}{ \tilde{\Psi}(s;-x)} = e^{\eta x}\quad,
 \label{eq:laplaceft}
\end{eqnarray}
which implies the fluctuation relation~(\ref{eq:fp}). 

We now  derive the fluctuation relation for the first-passage time distribution $\Phi(t;x)$ of the two boundary scenario,  given by Eq. (9) in the Main Text. 

The probability that the process $X$ reaches the boundary at $x$ having crossed $-x$ at least once before equals the probability that $X$ reaches $-x$ at $t'$ and then reaches unconditionally the barrier at $x$ in the time interval $t-t'$ summed over all mutually exclusive events over $t-t'$:
\begin{equation}
 \Omega(t;x) =  \int^t_0 \text{d}t'\Phi(t';-x)\Psi(t-t';2x)\quad.
 \label{eq:omegaa}
\end{equation}
Using Eq.~\eqref{eq:decomposition} and Eq.~\eqref{eq:omegaa}, $ \Psi (t;x) $   equals to
\begin{eqnarray}
 \Psi (t;x) &=& \Phi(t;x) +  \Omega(t;x) \nonumber\\
&=& \Phi(t;x) + \int^t_0 \text{d}t'\Phi(t';-x)\Psi(t-t';2x)\quad.\nonumber
\label{eq:FPTpm}
\end{eqnarray}
Taking Laplace transforms in both sides we obtain
\begin{equation}
 \tilde{\Psi}(s;x) =    \tilde{\Phi}(s;x) + \tilde{\Phi}(s;-x)\tilde\Psi(s;2x) \quad.
\end{equation}
We thus have that 
\begin{eqnarray}
\frac{ \tilde{\Psi}(s;x)}{ \tilde{\Psi}(s;-x) } &=&  \frac{\frac{ \tilde{\Phi}(s;x)}{ \tilde{\Phi}(s;-x) } + \tilde\Psi(s;2x)}{1 + \frac{ \tilde{\Phi}(s;x)}{ \tilde{\Phi}(s;-x) }\tilde\Psi(s;-2x) }\quad.
\end{eqnarray}
Solving for the ratio $ \tilde{\Phi}(s;x) / \tilde{\Phi}(s;-x) $ and using the fluctuation theorem in $\tilde{\Psi}$ given by Eq.~\eqref{eq:laplaceft} we obtain 
\begin{eqnarray}
\frac{ \tilde{\Phi}(s;x)}{ \tilde{\Phi}(s;-x) }  &=& \frac{e^{\eta x} - \tilde\Psi(s;2x) }{1 - e^{\eta x}  \tilde\Psi(s;-2x)  } 
\nonumber \\ 
 &=& e^{\eta x} \left(\frac{1-e^{-\eta x}\tilde\Psi(s;2x)}{1 - e^{\eta x}  \tilde\Psi(s;-2x)}\right) 
\nonumber \\ 
 &=&  e^{\eta x}\quad. \label{eq:ftfpo}
\end{eqnarray}
Equation~\eqref{eq:ftfpo} expresses a fluctuation theorem for the Laplace transform of the first-passage distribution $\Phi$:
\begin{equation}
\frac{ \tilde{\Phi}(s;x)}{ \tilde{\Phi}(s;-x) } = e^{\eta x}\quad,
\end{equation}
which yields equivalently
\begin{equation}
\frac{ {\Phi}(t;x)}{ {\Phi}(t;-x) } = e^{\eta x}\quad.
\label{eq:FTLT}
\end{equation}
Equation \eqref{eq:FTLT} corresponds to Eq.~(9) in the Main Text for $x=\Delta S_{\rm tot}$ and $\eta=1/k$. 

Equation \eqref{eq:FTLT} holds for processes where the transition probability is translationally invariant and satisfies the fluctuation theorem given by~\eqref{eq:FT}. Fluctuation theorem (13) with $x=\Delta S_{\rm tot}$ for the fluctuations of entropy production is known to hold for a wide class of processes including stochastic diffusive processes~\cite{kurchan1998fluctuation,lebowitz1999gallavotti,seifert2005entropy}, chaotic systems~\cite{gallavotti1995dynamical} and sheared fluids \cite{evans1993probability}.


\begin{thebibliography}{25}
\bibitem [{foo({\natexlab{a}})}]{footnote0}%
  \BibitemOpen
  \href@noop {} {\bibfield  {journal} {\bibinfo  {journal} {For simplicity, we
  only consider variables that are even under time reversal}\ }}\BibitemShut {NoStop}%
\bibitem [{foo({\natexlab{b}})}]{footnote3}%
  \BibitemOpen
  \href@noop {} {\bibfield  {journal} {\bibinfo  {journal} {Translational
  invariance implies that the transition probability $P(x|x' ; t,t')$ from
  state $x'$ to state $x$ only depends on the distance between states $P(x|x' ;
  t,t') = P(x-x';t,t')$}\ }}\BibitemShut {NoStop}%
\end{thebibliography}
\end{document}